\pdfoutput=1
\documentclass[12pt]{article}
\usepackage{amsmath,amssymb,emaxima}
\usepackage{geometry}
\usepackage{graphicx}

\newcommand{\tmem}[1]{{\em #1\/}}
\newcommand{\tmop}[1]{\ensuremath{\operatorname{#1}}}

\begin{document}

\title{Benchmarking the solar dynamo with Maxima\thanks{The text were
    processed with Emaxima (http://Maxima.sf.net)}}\author{Valery Pipin
(Institute Solar-Terrestrial Physics, Irkutsk)}\maketitle

Recently, Jouve et al \cite{jouve:08} published the paper that presents the numerical
benchmark for the solar dynamo models. Here, I would like to show a way how to
get it with help of computer algebra system Maxima. This way was used
in \cite{PS:08} to test some new ideas in the large-scale stellar dynamos. What you need are the
latest version of Maxima-5.16.3 (preferable compiled against the fastest lisps
like sbcl or cmucl-sse2) and some knowledge of the global (spectral) methods
to solve the PDE eigenvalue problem. For the quite comprehensive introduction
to these methods please look at the book by John Boyd \cite{boyd}. The basic steps to
solve the problem are:

1. the mathematical formulation (equation+boundary conditions)

2. choice the basis function and project equations to the basis

3. find matrices (apply some integration procedure in case of Galerkin
method)

4. apply linear algebra

The whole consideration is divided for two cases. As the first case we explore
the largest free decay modes in the sphere which is submerged in vacuum.
In this problem the all dynamo effects are neglected. As the second case I test the $\alpha \Omega$
dynamo in the solar convection zone with the tachocline
included.

Lets consider the spherical geometry. 
The evolution of the axisymmetric large-scale magnetic field (LSMF),
$\left\langle \mathbf{B} \right\rangle = \mathbf{e_{\phi}} B +
\mathrm{\tmop{curl}} \left( \frac{A \mathbf{e_{\phi}}}{r \sin \theta}
\right)$, (where $r$ is radius, $\theta$- co-latitude, $\mathbf{e}_{\phi}$- the
unit azimuthal vector) in the turbulent media subjected to the differential
rotation in the spherical shell can be described with equations:
\begin{eqnarray}
  \frac{\partial B}{\partial t} & = & \frac{1}{r} \frac{\partial \left(
  \Omega, A \right)}{\partial \left( r, \theta \right)} + \frac{1}{r} \left(
  \frac{\partial \left( r \mathcal{E}^{\theta} \right)}{\partial r} -
  \frac{\partial \mathcal{E}^r}{\partial \theta} \right),  \label{eq:1}\\
  \frac{\partial A}{\partial t} & = & r \sin \theta \mathcal{E}^{\phi}, 
  \label{eq:2}
\end{eqnarray}
In equations above, the turbulent contribution is expressed through the
components of the mean electromotive force (MEMF) $\mathcal{E} = \left\langle
\mathbf{u' \times b'} \right\rangle$, where $\mathbf{u}', \mathbf{b}'$ are the
small-scale fluctuated velocity and magnetic field respectively, $\Omega =
\Omega \left( r, \theta \right)$ -
the given angular velocity distribution.
For the sake of simplicity we restrict consideration to the case of $\alpha
\Omega$ dynamo 
with isotropic turbulent diffusion. Hence, we have
\begin{eqnarray}
  \mathcal{E}_r & = & - \frac{\eta_T}{r \sin \theta} \frac{\partial \sin
  \theta B}{\partial \theta}  \label{eq:er}\\
  \mathcal{E}_{\theta} & = & \frac{\eta_T}{r} \frac{\partial rB}{\partial r} 
  \label{eq:et}\\
  \begin{array}{l}
    r \sin \theta \mathcal{E}_{\phi}
  \end{array} & = & \left. \eta_T \left( \frac{\partial^2 A}{\partial r^2} +
  \frac{\sin \theta}{r^2} \frac{\partial}{\partial \theta} \frac{1}{\sin
  \theta} \frac{\partial A}{\partial \theta} \right) \right. + \hat{\alpha}
  \left( r, \theta \right) \eta_T G Br \sin 2 \theta 
\end{eqnarray}
where $\eta_T$- turbulent diffusion, $\hat{\alpha} \left( r, \theta \right)$ -
dimensionless function to model the $\alpha$ effect, \ $G = \nabla \log \rho$
- stratification parameter. We adopt the model parameters given in \cite{jouve:08}.
The boundary conditions are $\displaystyle{\frac{\partial rB}{\partial r} =
0, A = 0}$ - at the bottom and vacuum conditions - 
at the top of convection zone. For the computation all the equations
are written in dimensionless form with
new radial coordinate $x = r / R_{\odot}$. Moreover, to project equations on
the basis function we use the coordinate transformation to the interval where
basis is orthogonal.

As the first step we consider the solutions for the largest free-decay modes.
This intends to test the accuracy of the boundary conditions implementation procedure
and the speed of convergence. We neglect all the dynamo effects in
(\ref{eq:1}, \ref{eq:2}) and return to equations:
\begin{eqnarray}
  \frac{\partial B}{\partial t} & = & \frac{1}{x} \frac{\partial^2 \left( xB
  \right)}{\partial x^2} + \frac{1}{x^2} \frac{\partial}{\partial \theta}
  \frac{1}{\sin \theta} \frac{\partial \left( \sin \theta B \right)}{\partial
  \theta},  \label{A:eq:1}\\
  \frac{\partial A}{\partial t} & = & \frac{\partial^2 A}{\partial x^2} +
  \frac{\sin \theta}{x^2} \frac{\partial}{\partial \theta} \frac{1}{\sin
  \theta} \frac{\partial A}{\partial \theta},  \label{A:eq:2}
\end{eqnarray}
where for the sake of simplicity we have assumed that magnetic diffusivity is
constant over the depths and equations are written in dimensionless form.
Lets consider the integration domain on the radial coordinate to be $x \in \left[
0, 1 \right]$ 
and for latitude - from pole to pole. Having the regular
conditions at the origin and vacuum boundary conditions at the top, we
can get the analytical solutions of (\ref{A:eq:1}, \ref{A:eq:2}).
They can be expressed via the spherical Bessel functions. For the sake
of simplicity we restrict ourselves to solutions for the largest modes. 
We decompose the eigenmodes as follows
\begin{eqnarray}
  B \left( x, \theta, t \right) & = & e^{\lambda t} \sum_n \sum_m \hat{b}^{nm}
  S_n^{(B)} \left( x \right) P_m^1 \left( \cos \left( \theta \right) \right), 
  \label{N: eq:B-dec}\\
  A \left( x, \theta, t \right) & = & e^{\lambda t} \sum_n \sum_m \hat{a}^{nm}
  \sin \left( \theta \right) S_{nm}^{(A)} \left( x \right) P_m^1 \left( \cos
  \left( \theta \right) \right),  \label{N: eq:A-dec}
\end{eqnarray}
where $P_m^1$ are the associated Legendre polynomials and other
functions were found via basis recombination of the Legendre polynomials,
namely,
\begin{eqnarray}
  S_n^{(B)} \left( x \right) & = & x \left( P_{2 n + 1} \left( x \right) - P_1
  \left( x \right) \right),  \label{N: eq:B-dec1}\\
  S_{nm}^{(A)} \left( x \right) & = & x \left( P_{2 n + 1} \left( x \right) -
  P_1 \left( x \right) \frac{\left( \left( 2 n + 1 \right) \left( 2 n + 2
  \right) + 2 \left( m + 1 \right) \right)}{\left( 2 m + 4 \right)} \right), 
  \label{N: eq:A-dec1}
\end{eqnarray}
{\tmem{Here, each element of basis satisfies the boundary conditions
individually}}. For the largest decay modes 
we have $S_1^{(B)} \left( x
\right) \varpropto j_1 \left( \alpha x \right)$, $\lambda = -
\alpha^2$($\alpha \approx 4.4934$) and $S_1^{(A)} \left( x \right) \varpropto
j_1 \left( \pi x \right)$, $\lambda = - \pi^2$. Similar to
{\cite{liv-jac:05}}, the eigenvectors are scaled as $S_1^{(B)} \left( .5
\right) = 1$ and $S_1^{(A)} \left( 1 \right) = 1$. 
The errors are measured as $E \left( \lambda \right) = \left|
  \lambda_{true} - \lambda_{num} \right|$ and 
$E \left( \mathbf{B} \right) = \int_0^1 \left| \mathbf{B}_{true} -
\mathbf{B}_{num} \right|^2 dx$. The results of benchmark are given in the
Table \ref{tab1}.
\begin{center}
\begin{table}
\begin{centering}
\begin{tabular}{|c|c|c|}
\hline 
N & $E\left(\lambda\right),$B & $E$(B)\tabularnewline
\hline 
3 & 3.83e-5 & 6.849e-4\tabularnewline
\hline
\hline 
4 & 9.984e-8 & 4.365e-9\tabularnewline
\hline 
5 & 1.207e-10 & 2.497e-12\tabularnewline
\hline 
6 & 8.707e-14 & 9.068e-16\tabularnewline
\hline 
7 & 1.77e-14 & 4.80e-19\tabularnewline
\hline 
8 & 5.32e-15 & 6.263e-23\tabularnewline
\hline
\end{tabular}\begin{tabular}{|c|c|}
\hline 
$E\left(\lambda\right),$A & $E$(A)\tabularnewline
\hline 
1.08e-7 & 3.98e-9\tabularnewline
\hline
\hline 
5.651e-11 & 1.255e-12\tabularnewline
\hline 
9.68e-14 & 2.142e-16\tabularnewline
\hline 
7.72e-14 & 2.136e-20\tabularnewline
\hline 
1.38e-14 & 1.325e-24\tabularnewline
\hline 
\multicolumn{1}{|c||}{1.90e-14} & 4.427e-29\tabularnewline
\hline
\end{tabular}
\par\end{centering}
\caption{Convergence of the decay modes to the larges mode of analytic solution
of (\ref{A:eq:1},\ref{A:eq:2}). N is the number of modes in radial
basis. \label{tab1}}
\end{table}
\end{center}

Next benchmark is related to one given in the paper \cite{jouve:08}.
The detail of the model can be found in that paper. We consider the
results for test B, which is for $\alpha\Omega$ dynamo with external
vacuum boundary conditions and jump othe magnetic diffusivity below
the botom of convection zone. For the magnetic fields we have used
the following set of the modes:
\begin{eqnarray}
S_{n}^{(B)}\left(x\right) & = & \left(P_{n-1}\left(x\right)
-\frac{13\left(2n+1\right)P_{n}\left(x\right)}{13n^{2}+26n+6}
-\frac{\left(13n^{2}-7\right)P_{n+1}\left(x\right)}{13n^{2}+26n+6}\right),\label{A:eq:B-dec1}\\
S_{nm}^{(A)}\left(x\right) & = & x\left(P_{n}\left(x\right)
+\frac{\left(2n+3\right)P_{n+1}\left(x\right)}{n^{2}+4n+2m/f+4}
-\frac{\left(n^{2}+4n+2m/f+1\right)P_{n+2}\left(x\right)}{n^{2}+4n+2m/f+4}\right),
\label{A: eq:A-dec1}\end{eqnarray}
where factor $f={\displaystyle \frac{2}{1-x_{i}}}$ appears due the
coordinate transformation $\left[x_{i},1\right]\rightarrow\left[-1,1\right]$
and $x_{i}=0.65$ is bottom of integration domain.

\begin{figure}
\begin{centering}
\includegraphics[width=0.8\textwidth]{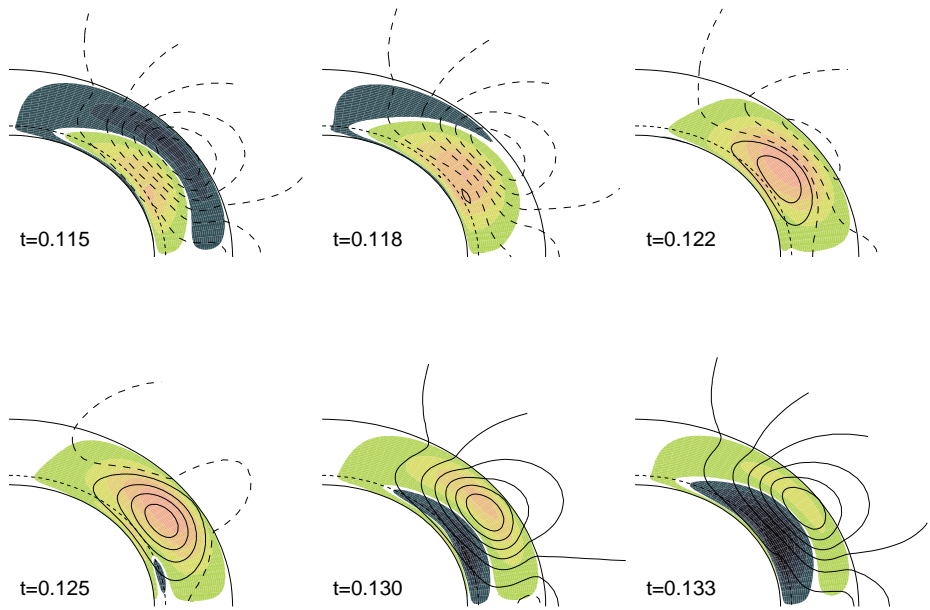}
\par\end{centering}

\caption{Model B of \cite{jouve:08}. Snapshots of evolution of
  torioidal magnetic fileds (shadowed density plot) and poloidal
  fields (shown with streamlines) over the half of cycle. \label{modB}}
\end{figure}

The results are shown at the table 2 and Figure \ref{modB}.
\begin{table}
\begin{centering}
\begin{tabular}{|c|c|c|}
\hline 
Resolution & $C_{\alpha}^{crit}$ & $\omega$\tabularnewline
\hline
\hline 
8x8 & .443 & 180.5\tabularnewline
\hline 
10x10 & .4175 & 175.1\tabularnewline
\hline 
12x12 & .4095 & 172.2\tabularnewline
\hline 
13x13 & .411 & 172.4\tabularnewline
\hline 
14x14 & .4122 & 172.7\tabularnewline
\hline 
16x16 & .4125 & 172.9\tabularnewline
\hline
\end{tabular}
\par\end{centering}
\caption{Model B of \cite{jouve:08}. Resolution $N\times M$ means that
N modes is used in decomposition on radius and M - on latitude.}
\end{table}
Note, that our results are quite close to those in \cite{jouve:08}. The
small difference can be attributed to the difference in the method  of
solution of the problem. 

\section*{The largest decay mode of the diffusion operator}
Here, I give some details of the above computation within Maxima. The
strongest side  of the computer algebra system is that 
most of the computational work is done exactly. For example, all the
derivatives in (\ref{A:eq:1}, \ref{A:eq:2}) are computed directly by
substitution of (\ref{N: eq:B-dec}, \ref{N: eq:A-dec}) to (\ref{A:eq:1},
\ref{A:eq:2}). The  integration  over
space domain has to be done to proceed with the Galerkin method.
This is carried out with help of the Gauss-Legendre
procedure. The procedure is based on the so-called "collocation points
and weights" method. The collocation
points are the set of zeros of the Legendre polynomial
of the order $> N$ where $N$ is the number of the greatest mode among $n = 1, \ldots N$
in (\ref{N: eq:B-dec}, \ref{N: eq:A-dec}). The collocation points
and weights can be found via subprogram "pseudp.mac". It is given in
Appendix.  The maxima session is started with the general definitions:
\begin{maxima}[]
kill(all)$
batchload("pseudp.mac")$
(xb:0.,xe:1.)$
bftorat:true$ 
float2bf:true$
ratprint:false$
fpprec:64$ 
ratepsilon:1.e-32$
chn(x):=x$
realGr(L1,L2):=(if realpart(L1) >= realpart(L2) then true else false)$
\end{maxima}
Here, we find the collocation points and weights on radius,
\begin{maxima}[]
(np:5,Nch:4*np,xchwh:gaulegP(-1,1,Nch,1), xch:xchwh[1],wh:xchwh[2])$  
\end{maxima}
and on latitude,
\begin{maxima}[]
(NT:6,NTT:6*NT,mut_w:gaulegP(-1,1,NTT,1),mut:mut_w[1],w:mut_w[2])$    
\end{maxima}
Next, we define the indices for the basis functions and the parities
for the toroidal and poloidal eigenmodes,
\begin{maxima}[]
Nch:length(xch)$
NTT:length(mut)$
for m:1 thru NT do(   for n:1 thru np do(
        ki:np*(m-1)+n,ni[ki]:n,mi[ki]:m))$
ni:makelist(ni[i],i,1,NT*np)$
mi:makelist(mi[i],i,1,NT*np)$
NN:NT*np$
parity_b:0$
parity_a:0$
\end{maxima}
Note, that only a half of $[-1,1]$ intervals is used in
computations. This leads to the computational economy and to increase
the spectral resolution of the code.
The following are the part of the code which defines the basis functions
on radius ($S^{(B)}_n\rightarrow chtb_n$, $S^{(A)}_{nm}\rightarrow chta_{nm}$ ) and latitude,
\begin{maxima}[]
kill(chtb1,chtb)$
chtb[n](x):=x*(legendre_p(1,x)-legendre_p(2*n+1,x))$ 
makelist(taylor(chtb[i](x),x,0,1),i,1,np)$  
plot2d(makelist(chtb[i](x),i,1,np),[x,-1,1])$
kill(chta00)$
chta00(n,l,x):=x*(legendre_p(2*n+1,x)
  -((2*n+1)*(2*n+2)+2*l+2)*legendre_p(1,x)/(2*l+4))$
chta[n,l](x):=chta00(n,2*l-1,x),       
kill(Pl10,Pla,Plb)$
Pl10(n,x):=expand((sqrt((2*n+1)/(2*n*(n+1)))*assoc_legendre_p(n,1,x)))$  
Pla[n](x):=block(if parity_a=0 then bfloat(Pl10(2*n-1,x)) else bfloat(Pl10(2*n,x)) ) $  
Plb[n](x):=block(if parity_b=0 then bfloat(Pl10(2*n-1,x)) else bfloat(Pl10(2*n,x)) ) $    
\end{maxima}
As you see, we use odd or even associated Legendre polynomial in
respective of the parity choice. To accelerate the integration
procedure we calculate the matrices of basis functions over the sets
of collocation points,
\begin{maxima}[]
remarray(PLB,PLA)$
kill(PLB1,PLA1)$
PLB[n,i]:=bfloat(Plb[n](mut[i]))$
PLA[n,i]:=bfloat(Pla[n](mut[i]))$
PLA1[n,i]:=bfloat(Pla[n](mut[i])*w[i])$
PLB1[n,i]:=bfloat(Plb[n](mut[i])*w[i])$
CHTB[n,j]:=bfloat(chtb[n](xch[j]))$
CHTA[n,m,j]:=bfloat(chta[n,m](xch[j]))$
makelist(makelist(makelist(CHTA[n,m,j],n,1,np),m,1,NT),j,1,Nch)$
CHTB1[n,j]:=bfloat(chtb[n](xch[j])*wh[j])$
CHTA1[n,m,j]:=bfloat(chta[n,m](xch[j])*wh[j])$
makelist(makelist(makelist(CHTA1[n,m,j],n,1,np),m,1,NT),j,1,Nch)$
genmatrix(PLB,NT,NTT)$
genmatrix(PLA,NT,NTT)$
genmatrix(PLA1,NT,NTT)$
genmatrix(PLB1,NT,NTT)$
genmatrix(PLB2,NT,NTT)$
genmatrix(CHTB,np,Nch)$
genmatrix(CHTB1,np,Nch)$
\end{maxima} 
Now we can to proceed to solution of the problem  (\ref{A:eq:1},
\ref{A:eq:2}). They can be treated separately. Firstly we define the
left parts of  (\ref{A:eq:1},\ref{A:eq:2}),
\begin{maxima}[]
kill(AA,BB)$
AA[ki,kj]:=bfloat(sum(CHTA1[ni[ki],mi[ki],j]*CHTA[ni[kj],mi[kj],j]*
    sum(PLA1[mi[ki],i]*PLA[mi[kj],i],i,1,NTT),j,1,Nch))$
BB[ki,kj]:=bfloat(sum(CHTB1[ni[ki],j]*CHTB[ni[kj],j]
    *sum(PLB1[mi[ki],i]*PLB[mi[kj],i],i,1,NTT),j,1,Nch))$
BB:genmatrix(BB,NN,NN)$
AA:genmatrix(AA,NN,NN)$
\end{maxima}
Note, the integration is just a product of sums over collocation
points. Now lets compute the right part of (\ref{A:eq:1}). The computations
are straightforward. The radial part is, 
\begin{maxima}[]
kill(CHTB_d2)$
CHTB_d2[n,j]:=bfloat(subst([x=xch[j]],
    expand((diff(diff(chtb[n](x),x),x) ))))$
genmatrix(CHTB_d2,np,Nch)$
kill(et_d)$
et_d[ki,kj]:=bfloat(sum(CHTB_d2[ni[kj],j]*CHTB1[ni[ki],j],j,1,Nch)
  *sum(PLB[mi[kj],i]*PLB1[mi[ki],i],i,1,NTT))$
et_d:genmatrix(et_d,NN,NN)$
\end{maxima}
The latitudinal part is computed as follows
\begin{maxima}[]
kill(PLB_d2,PLB_d2m,PLB_d1,PLB_d)$
PLB_d2m[n,i]:=bfloat(subst([mu=mut[i]],expand((sqrt(1-mu^2)
    *diff(diff(sqrt(1-mu^2)*Plb[n](mu),mu),mu)))))$
genmatrix(PLB_d2m,NT,NTT)$
kill(er_d)$
er_d[ki,kj]:=bfloat(sum(CHTB1[ni[ki],j]*CHTB[ni[kj],j]/chn(xch[j])^2,j,1,Nch)
    *sum(PLB1[mi[ki],i]*PLB_d2m[mi[kj],i],i,1,NTT))$
er_d:genmatrix(er_d,NN,NN)$
\end{maxima}
Invert the left part of  and solve the eigenvalue problem with help of
lapack
\begin{maxima}[]
BMI:invert_by_lu(BB)$
realGr0(L1):=(if realpart(L1) >=0 then true else false)$
load(lapack)$
MT:BMI.(er_d+et_d)$
lamb:(dgeev(MT,true))$
\maximaoutput
...
;; Loading #P"/usr/share/maxima/5.16.3/share/lapack/lapack/binary-cmucl/dtrevc.sse2f".
\end{maxima}
Now, lets sort the eigenvalues,
\begin{maxima}[]
lmb:sort(flatten((col(lamb,1))[1]),realGr);
\maximaoutput
       [- 20.19072855751148, - 48.83125052872499, - 59.68036461428608, 
- 87.54250816953108, - 108.8085726200047, - 119.8731393320569, 
- 135.9006458324267, - 169.2937978826357, - 193.7765731289586, 
- 203.8596788530381, - 238.1199364033153, - 241.2729384729356, 
- 260.8709651925637, - 277.930972796181, - 320.6253098224506, 
- 328.0254017082749, - 373.0041226108443, - 410.8018157235754, 
- 505.412713312208, - 577.4557504236772, - 665.4381736491396, 
- 809.3382250971157, - 851.3024977594987, - 882.9750962899574, 
- 932.6139588501989, - 970.2769707279747, - 1049.716092377411, 
- 1259.650734653339, - 1830.07144488619, - 2585.672953057317]
\end{maxima}
The relevant spherical Bessel functions of the problem are 
\begin{maxima}[]
jn(n,x):=spherical_bessel_j(n,x)$
jnR(n,x):=x*spherical_bessel_j(n,x)$
\end{maxima}
Next we compute the first root of $j_n(x)$ and compare it  with the first eigenvalue,
\begin{maxima}[]
a1:find_root(jn(1,x),x,4,5)$
sqrt(abs(first(lmb)))-a1;
\maximaoutput
                             1.207158817351228e-10
\end{maxima}
Similarly, we solve (\ref{A:eq:2}). At the first, we invert the left
part of it
\begin{maxima}[]
AMI:invert_by_lu(AA)$
\end{maxima}
Then calculate the right part and apply lapack solver,
\begin{maxima}[]
kill(CHTA_d2,PLA_d2)$
CHTA_d2[n,m,j]:=subst([x=xch[j]],(diff(diff(chta[n,m](x),x),x)))$
makelist(makelist(makelist(CHTA_d2[n,m,j],n,1,np),m,1,NT),j,1,Nch)$
PLA_d2[n,i]:=bfloat(subst([mu=mut[i]],(sqrt(1-mu^2)
    *diff(diff(sqrt(1-mu^2)*Pla[n](mu),mu),mu))))$
genmatrix(PLA_d2,NT,NTT)$
kill(ef_d)$
ef_d[ki,kj]:=bfloat(sum(CHTA_d2[ni[kj],mi[kj],j]
    *CHTA1[ni[ki],mi[ki],j],j,1,Nch)*
   sum(PLA1[mi[ki],i]*PLA[mi[kj],i],i,1,NTT)
  +sum(CHTA[ni[kj],mi[kj],j]*CHTA1[ni[ki],mi[ki],j]
    /chn(xch[j])^2,j,1,Nch)*
    sum(PLA1[mi[ki],i]*PLA_d2[mi[kj],i],i,1,NTT))$
ef_d:genmatrix(ef_d,NN,NN)$
MT:(AMI.ef_d)$
lamb:(dgeev(MT,true))$
\end{maxima}
Then, we sort the eigenvalues and compare the largest eigen mode with the
first zero of $J_{1/2}(x)$, 
\begin{maxima}[]
lmb1:sort(flatten((col(lamb,1))[1]),realGr)$
float(sqrt(abs(first(lmb1)))-
\maximaoutput
                            - 9.681144774731365e-14
\end{maxima}
To compare the eigenfunctions we have to apply the scaling  $S_1^{(B)} \left( .5
\right) = 1$ and $S_1^{(A)} \left( 1 \right) = 1$. 

The solution of dynamo problem is obtained in essentially the same
way. I will put the code for it somewhere in the public place.


\section*{Appendix}
The subroutine to find the collocation points and weghts for the
polynomial order of n in interval $[x1,x2]$. The Newton method is applied,
\begin{maxima}[]
load(orthopoly)$
orthopoly_returns_intervals:false$
fpprec:32$
float2bf:true$
ratprint:false$
load("linearalgebra")$
gaulegP(x1,x2,n):= block([xm,wm,A0,i,mm,pp,xm,xl,z,z1,numer,eps],
kill(xt,w), eps:1.e-16,fpprec:32,  
(if oddp(n) then mm:(n+1)/2 else mm:n/2),  
xm:0.5*(x2+x1), xl:0.5*(x2-x1),  
for i: 1 while i <= mm do (  
z:bfloat(cos(
do (pp:bfloat(n*(z*legendre_p(n,z)-legendre_p(n-1,z))/(z*z-1.0)),  
z1:z,z:bfloat(z1-legendre_p(n,z)/pp),  
if abs(z-z1) < eps  
then return(z1)),  
xt[i]:xm-xl*z1,  
xt[n+1-i]:xm+xl*z1),  
A:bfloat(genmatrix(lambda([i,j],legendre_p(i-1,xt[j])),n,n)),  
RH:bfloat(transpose(matrix(  
makelist((if i=1 then defint(legendre_p(0,x),x,-1,1) else 0),i,1,n)))),  
w: bfloat(invert_by_lu(A) . RH),
return([makelist(xt[i],i,1,n),flatten(makelist(w[i],i,1,n))]))$  
\end{maxima}
\end{document}